# The potential for photosynthesis in hydrothermal vents: a new avenue for life in the Universe?

[1]Noel Perez, [1]Rolando Cardenas, [1]Osmel Martin, [2] Leiva-Mora Michel

[1] Planetary Science Lab. Department of Physics, Universidad Central "Marta Abreu" de Las Villas, Santa Clara, Cuba.     Phone: 53 42 281109. Fax: 53 42 281130.

[2] Institute for Plant Biotechnology. Universidad Central "Marta Abreu" de Las Villas, Santa Clara, Cuba.  Phone: 53 42 281257. Fax: 53 42 281329

E-mail:     noelpd@uclv.edu.cu;      rcardenas@uclv.edu.cu;      osmel@uclv.edu.cu; michel@ibp.co.cu

**Abstract:** We perform a quantitative assessment for the potential for photosynthesis in hydrothermal vents in the deep ocean. The photosynthetically active radiation in this case is from geothermal origin: the infrared thermal radiation emitted by hot water, at temperatures ranging from 473 up to 673 K. We find that at these temperatures the photosynthetic potential is rather low in these ecosystems for most known species. However, species which a very high efficiency in the use of light and which could use infrared photons till 1300nm, could achieve good rates of photosynthesis in hydrothermal vents. These organisms might also thrive in deep hydrothermal vents in other planetary bodies, such as one of the more astrobiologically promising Jupiter satellites: Europa.

## I Introducction

Light energy from the Sun drives photosynthesis to provide the primary source of nearly all of the organic carbon that supports life on Earth (Blankenship 2002). An alternative energy source can be found in hydrothermal vents, such as black smokers located far below the photic zone in the oceans, where unusual microbial and invertebrate populations exist on organic material  from $CO_2$ reduction by chemotrophic bacteria that oxidize inorganic compounds (Van Dover 2000).  Hydrothermal vents may resemble the environment in which life evolved (Martin et al. 2003, Simoncini et al 2011), and the discovery of geothermal light at otherwise dark deep-sea vents led to the suggestion that such light may have provided a selective advantage for the evolution of photosynthesis from a chemotrophic microbial ancestor that used light-sensing molecules for hototaxis toward nutrients associated with geothermal light (Van Dover et al. 1996, Nisbet et al. 1995).
A bacterium that appears to use light as an auxiliary source of energy to supplement an otherwise chemotrophic metabolism was isolated from the general vicinity of a deep-sea hydrothermal vent (Yurkov et al. 1999, Beatty 2002, Beatty et al 2005). The discovery of such an organism in this environment would indicate that volcanic or geothermal light is harvested to drive photosynthetic reactions in the absence of light from the Sun. The possibility of geothermal light-driven photosynthesis on Earth relates to speculations about the existence of extraterrestrial life on planets and moons far from the Sun in the Solar System (Chyba 2001) and, conceivably, in other galaxies.
However, largely due to the high costs of deep sea explorations, hydrothermal vents near submarine volcanoes are far from being thoroughly studied. Thus, in this work we apply a mathematical model of photosynthesis to theoretically assess the photosynthetic potential in deep sea hydrothermal vents.

**II Materials and methods**

First we consider a source of geothermal photons, emitting the same flux as the TY black smoker of the East Pacific Rise, where a green sulphur bacterium (GSB1) was captured and studied (Beatty et al 2005). At the temperature of TY orifice (643 K) the photon flux in the visible part of the spectrum is very small compared to the infrared one, and thus it was neglected. The cells of GSB1 absorbed infrared photons mainly in the range (700 – 800) nm, so we only considered this light, with flux (radiance) of $10^8$ photons/cm$^2$.s.sr. For the first calculations, we consider a spherical source of geothermal photons, thus to above radiance to irradiance we multiplied by the solid angle subtended by a sphere ($4\pi$ sr). The irradiance $E(r)$ at a distance $r$ from the source is given by:

$$E(r) = \left(\frac{R}{r}\right)^2 E(R) \qquad (1)$$

where R is the radius of the source and $E(R)$ is the irradiance leaving the source's surface.

In a second group of calculations, we considered the fact that the water that surrounds a black smoker is also hot and will emit infrared photons too; therefore it is of interest to consider a distributed source. This is more realistic than a rather localized spherical source. For this case we assume a grey body approximation for the emission of photons, which means that the emissivity ($\varepsilon$) is considered independent of wavelength. For hot water, emissivity is often taken to be 0.95 and the emitted spectral irradiances $E(\lambda,T)$ are expressed by:

$$E(\lambda,T) = \varepsilon \cdot E_{bb}(\lambda,T) \qquad (2)$$

In the above expression $E_{bb}(\lambda,T)$ are the spectral irradiances of the blackbody at the same temperature $T$, given by the Planck's radiation law:

$$E(\lambda,T) = \frac{2\pi hc^2}{\lambda^5} \cdot \frac{1}{\exp(hc/\lambda kT) - 1} \qquad (3)$$

where $c, h, k$ are the light speed, Planck's and Boltzmann's constants respectively, $\lambda$ is the wavelength and $T$ the water's temperature.
Total irradiances $E_{PAR}(T)$ at temperature $T$, for the case of photosynthetically active radiation (PAR), are calculated by:

$$E_{PAR}(T) = \sum_{\lambda_i}^{\lambda_f} E(\lambda, T) \cdot \Delta\lambda \qquad (4)$$

being $\lambda_i$ and $\lambda_f$ the extreme wavelengths of the PAR band. For this second set of calculations, first we used the same range as with the spherical source ($\lambda_i = 700$ nm and $\lambda_f = 800$ nm), for the sake of comparison. Later we extended the infrared range, using $\lambda_f = 1100$ nm and $\lambda_f = 1300$ nm. $\Delta\lambda$ is the width of the intervals between the wavelengths. In our case, $\Delta\lambda = 1$ nm.

To account for the photosynthesis rates $P$ (normalised to the maximum rates $P_S$), we used the so-called $E$ photosynthesis model for phytoplankton, which assumes good repair capabilities against damage caused by ultraviolet radiation (Fritz et al. 2008):

$$\frac{P}{P_s} = \frac{1 - e^{E_{PAR}(T)/E_S}}{1 + E_{inh}^*(T)} \qquad (5)$$

where $E_{inh}^*(T)$ means that is a biologically effective irradiance, as the physical one was convolved (weighted) with a biological action spectrum.

Ultraviolet radiation coming from the Sun (or another parent star) would be absorbed in the first tens of meters of the water column. Therefore, at deep sea hydrothermal vents, this radiation would not produce the inhibitory effect on photosynthesis which usually produces on surface aquatic ecosystems in Earth. Thus, substituting $E_{inh}^*(T) = 0$, the above model results in:

$$\frac{P}{P_S} = 1 - e^{E_{PAR}(T)/E_S} \qquad (6)$$

where $E_S$ is a parameter indicating the efficiency of the species in the use of PAR, inversely proportional to the quantum yield of photosynthesis: the smaller $E_S$, the more efficient the species is. It represents the irradiance at which 63% of maximum photosynthesis $P_S$ is achieved. We sampled $E_S$ in a range, spanning from 5 W/m² up to 100 W/m², considering that most species on Earth would respond inside this range. However, we also considered the exceptional capacities of green sulphur bacteria in using even some part of the infrared band, and then explored down to the range 0.5 W/m² up to 2.5 W/m² (Pringault et al. 1998). Indeed, visual inspection of Fig. 5 of the above reference suggests $E_S \sim 0,5$ W/m².

**III Results and discussion**

Figure 1 shows that for the spherical source, photosynthesis rates are very small, of the order $10^{-4}$.

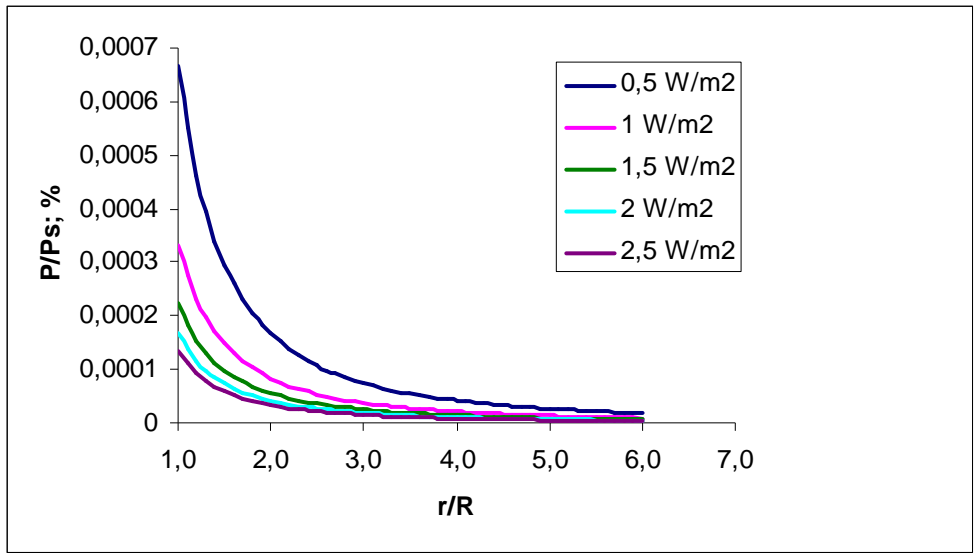

**Fig. 1** Relative photosynthesis rates for geothermal photons from a spherical source. The values of the parameter $E_S$ are indicated in the box.

Figure 2 shows a more realistic case, where it is considered that photons are emitted in all the surroundings of the black smoker: a distributed source. In this case, for temperatures near the one around TY black smoker (643 K), relative photosynthesis rates are of order $10^{-3}$.

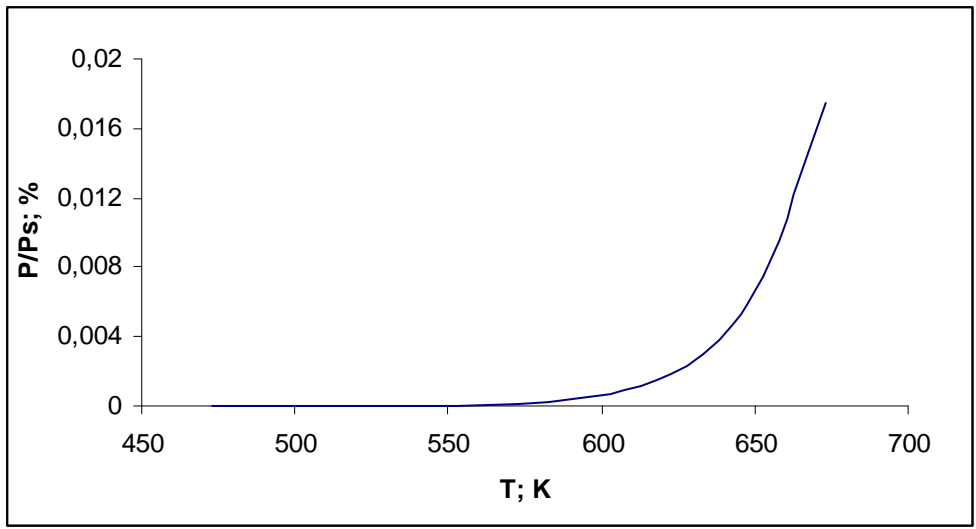

**Fig. 2** Relative photosynthesis rates for geothermal photons from a spherical source. We considered only the most efficient organisms ($E_S = 0,5$ W/m$^2$).

Above figures suggest that photosynthetic life in hydrothermal vents can only thrive if they are able to use a broader range of the infrared band. In Figure 3 we show the photosynthesis rates for organisms using infrared from 700 nm to 1100 nm, with an $E_S$ in the interval of 5W/m$^2$-100 W/m$^2$. At the temperature of black smoker TY, rates are of order $10^{-1}$ for efficient organisms.

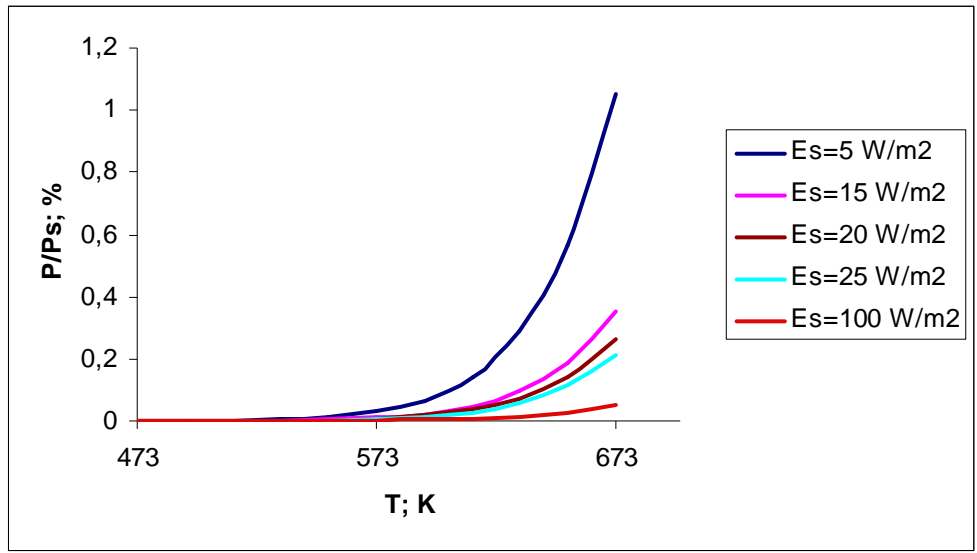

**Fig. 3** Relative photosynthesis rates. The values of the parameter $E_S$ are indicated in the box.

These results suggest that the photosynthetic organisms actually living in hydrothermal vents should use a larger (infrared) wavelength range and/or being more efficient using PAR, if they are supposed to be numerically abundant. Thus, we carried out other calculations increasing the PAR wavelength till to 1300 nm, while keeping $E_S$ from 5 W/m$^2$ to 100 W/m$^2$. Results are shown in Figure 4.

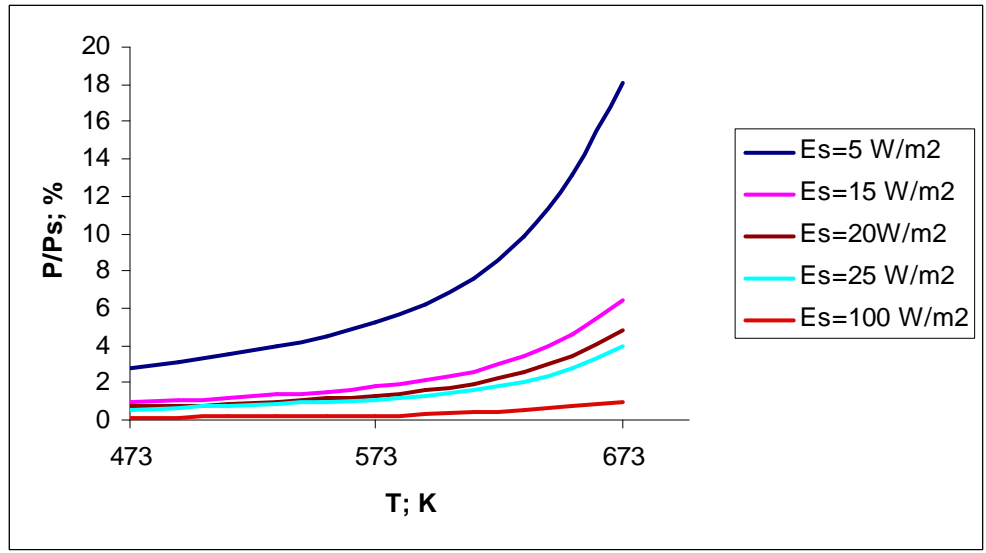

**Fig. 4** Relative photosynthesis rates. The values of the parameter $E_S$ are indicated in the box.

The photosynthesis rates are greater now, but still not high (no more than 18 % as maximum for the most efficient organism). For this reason, we made another study working with wavelengths between 700 nm and 1100 but with the $E_S$ parameter from 0.5 W/m$^2$ to 2.5 W/m$^2$, supposing more efficient organisms, such as green sulphur bacteria (Pringault et al 1998). Figure 5 shows the results.

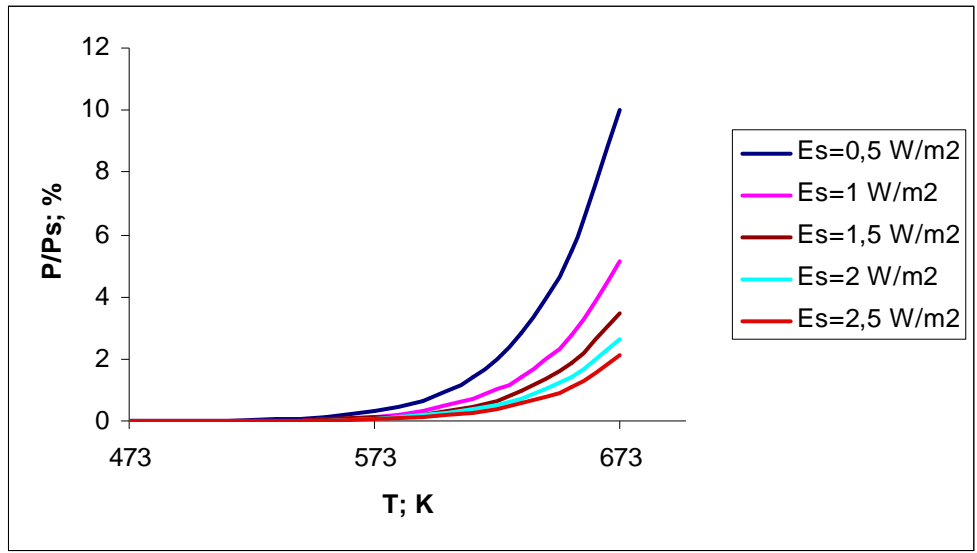

**Fig. 5** Relative photosynthesis rates. The values of the parameter $E_S$ are indicated in the box.

The photosynthesis rates again are not high, being of around 8% at temperatures of black smoker TY even for the more efficient organisms. Then, finally, we use a PAR wavelength range from 700 nm to 1300 nm and $E_S$ between 0.5 W/m$^2$ and 2.5 W/m$^2$. Results are shown in Figure 6.

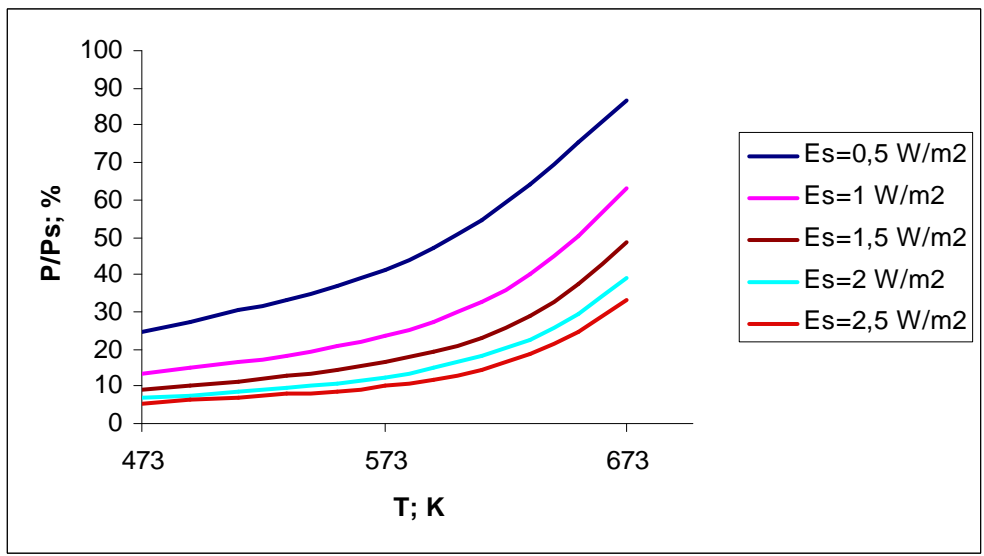

**Fig. 6** Relative photosynthesis rates. The values of the parameter $E_S$ are indicated in the box.

Now the photosynthesis rates are considerably higher, even comparable to those in the so called photic zone of surface waters on Earth. For the sake of comparison, we refer the reader to the works, by some of us, with photosynthesis in surface waters: (Avila, Cardenas & Martin 2012) and (Perez et al 2013).

**Conclusions**

Hydrothermal vents have the potential for hosting photosynthetic life using infrared radiation emitted by hot water. There is and advantage concerning surface waters: no inhibitory ultraviolet radiation. However, the extent to which photosynthesis will be actually performed; will depend on both the efficiency of the species using (infrared) photosynthetically active radiation, and the part of the infrared band that can be really used in photosynthesis. In this work we showed that very efficient organisms already known to use infrared radiation for photosynthesis (such as green sulphur bacteria) can have high photosynthesis rates provided they can use the infrared band up to 1300nm. As the deep sea vents on Earth are far from being well studied, we argue that there are possibilities for some organisms to thrive there using infrared light to photosynthesize. Of course, this also shows perspectives for such life forms in other planetary bodies potentially hosting deep sea vents, one example being Europa, one of the most popular moons of planet Jupiter.